\documentclass[pra,amsmath,amssymb,twocolumn,superscriptaddress]{revtex4-2}
 \usepackage{amsmath}
\usepackage{amssymb}
\usepackage{amstext}
\usepackage{amsfonts}
\usepackage{amsxtra}
\usepackage{bm}
\usepackage[usenames]{color}
\usepackage{grffile}
\usepackage{soul}
\usepackage{epstopdf}
\usepackage{upgreek}
\usepackage[dvipsnames]{xcolor}
\usepackage{graphicx}
\usepackage{marvosym}
\usepackage{wasysym}
\usepackage{physics}

\usepackage[colorlinks=true, letterpaper=true, pdfstartview=FitV,
linkcolor=blue, citecolor=blue, urlcolor=blue]{hyperref}

\usepackage[T1]{fontenc}

{%
\setlength{\fboxsep}{0pt}%
\setlength{\fboxrule}{1pt}%
}%
\begin{document}

\title{Magnetic soliton molecules in binary condensates}

\author{R. M. V. R\"{o}hrs}
\affiliation{
     Universit\"{a}t Innsbruck, Fakult\"{a}t f\"{u}r Mathematik, Informatik und Physik, Institut f\"{u}r Experimentalphysik, 6020 Innsbruck, Austria
}

\author{Chunlei Qu}
\affiliation{Department of Physics, Stevens Institute of Technology, Hoboken, NJ 07030, USA}
\affiliation{Center for Quantum Science and Engineering, Stevens Institute of Technology, Hoboken, NJ 07030, USA}

\author{R. N. Bisset}
\affiliation{
     Universit\"{a}t Innsbruck, Fakult\"{a}t f\"{u}r Mathematik, Informatik und Physik, Institut f\"{u}r Experimentalphysik, 6020 Innsbruck, Austria
}

\begin{abstract}
Two-component Bose-Einstein condensates in the miscible phase can support polarization solitary waves, known as magnetic solitons.
By calculating the interaction potential between two magnetic solitons, we elucidate the mechanisms and conditions for the formation of bound states---or {\it molecules}---and support these predictions with dynamical simulations.
We analytically determine the {\it dissociation energy} of bound states consisting of two oppositely polarized solitons and find good agreement with full numerical simulations.
Collisions between bound states---either with other bound states or with individual solitons---produce intriguing dynamics. Notably, collisions between a pair of bound states exhibit a dipole-like behavior.
We anticipate that such bound states, along with their rich collision dynamics, are within reach of current experimental capabilities.

\end{abstract}

\date{\today}
\maketitle

\section{Introduction}

Stabilized by a balance between dispersion and nonlinearity, solitons are localized collective excitations in elongated systems, relevant for a broad range of systems ranging from water in narrow channels \cite{Russell1885the,Ablowitz2011nonlinear} to optical fibers \cite{Kivshar2003optical}, plasmas \cite{Kono2010nonlinear} and conducting polymers \cite{Heeger1988solitons,Dauxois2006physics}.
Moreover, ultracold Bose-Einstein condensates (BECs) provide an especially appealing platform for the investigation of solitons, due to the clean and highly controllable experimental apparatuses available.
As classic examples, BECs with defocusing nonlinearities can support dark solitons \cite{Burger1999dark,denschlag2000generating,dutton2001observation,Anderson2001watching}, while bright solitons may form in the presence of focusing interactions \cite{strecker2002formation,khaykovich2002formation}.

Multi-component BECs allow one to investigate the rich physics offered by vector solitons, non-dispersive waves spanning multiple coupled components \footnote{It is worth noting that, although vector solitons may not always be true solitons in the strict mathematical sense, they possess key soliton characteristics, and we will refer to them as solitons for consistency with the literature.}.
For the case of two distinguishable components \cite{myatt1997production,hall1998dynamics,maddaloni2000collective,papp2008tunable,thalhammer2008double,papp2008tunable}---a {\it binary condensate}---experiments have observed dark-bright \cite{becker2008oscillations} and dark-antidark solitons \cite{hoefer2011dark,yan2012beating,danaila2016vector,Katsimiga2020,mossman2024observation}, as well as the closely related magnetic solitons \cite{farolifi2020observation,chai2020magnetic} (see related theory \cite{qu2016magnetic,qu2017magnetic}), which occur in balanced BEC mixtures close to the miscible-immiscible transition.

Soliton bound states---composite structures of multiple solitons---were predicted by \"Ohberg and Santos using pairs of magnetic solitons of opposite polarization \cite{Oehnberg2001dark}\footnote{In Ref.~{\cite{Oehnberg2001dark}}, magnetic solitons are referred to as kink-antikink solitons.} (also see more recent works \cite{charalampidis2016SO2,wang2021dark}). Closely related bound structures were later created in experiments by accelerating two overlapping condensates in opposite directions to create counterflows \cite{hoefer2011dark,yan2012beating}. 
Despite being predicted relatively early, much remains unknown about magnetic soliton bound states and their underlying physical mechanisms.

\begin{figure}[h!]
	\centering
	\includegraphics[width=0.45\textwidth]{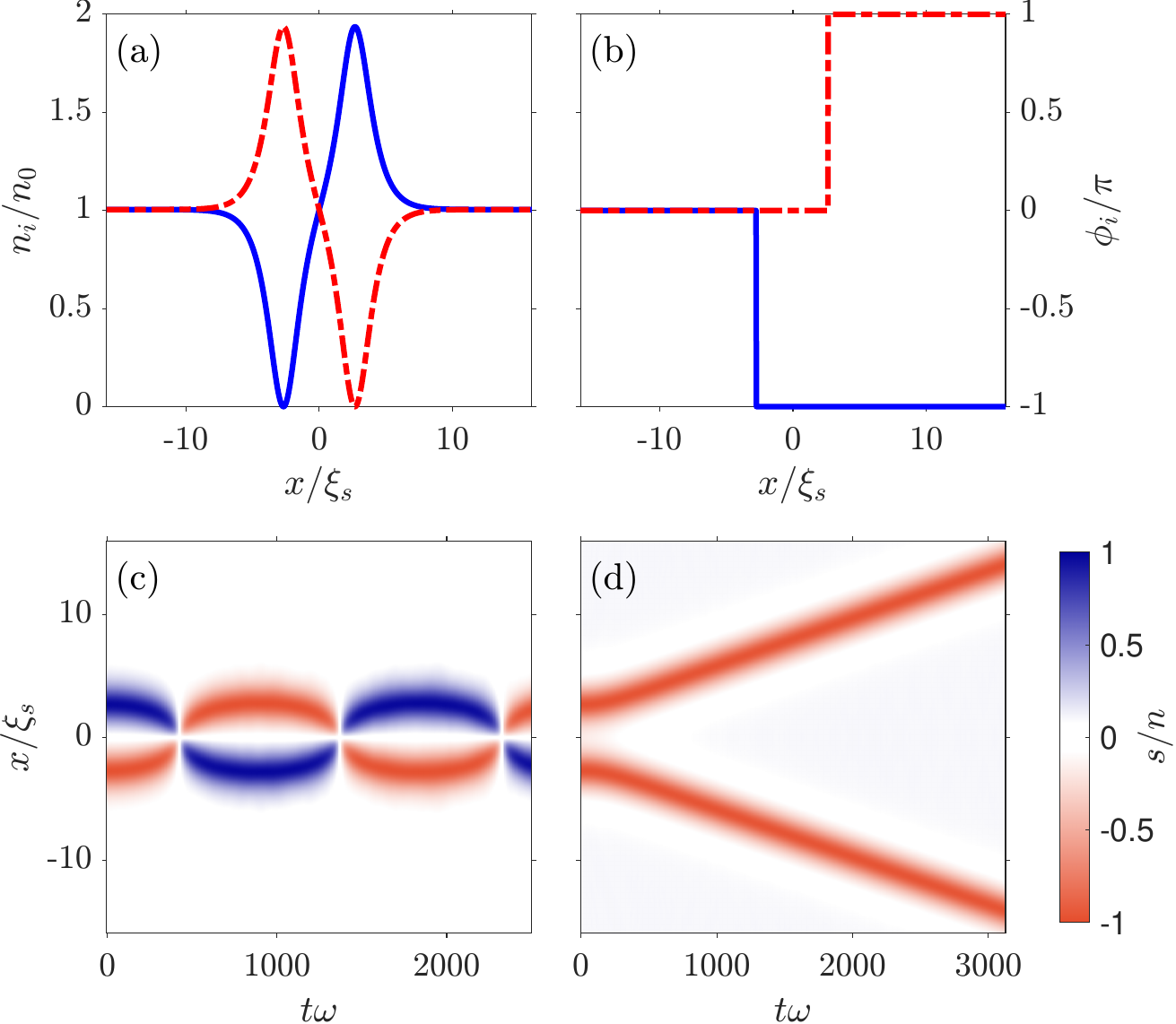}
	\caption{(a) Density and (b) wavefunction phase as a function of position for two zero-velocity magnetic solitons with opposite polarizations at $t=0$, calculated using Gross-Pitaevskii theory. Component 1 (2) is shown with a solid (dot-dashed) line.
(c) Spin density dynamics, $s=n_1-n_2$, starting from the above soliton pair, which forms an excited bound state.
(d) Dynamics of a like-polarization pair, illustrating the contrasting case in which the solitons repel and are unbound. 
	 }
	\label{Fig1:DensPhase}
\end{figure}

In this paper, we theoretically uncover the properties and underlying mechanisms of bound states formed by magnetic solitons in two-component BECs.
To build intuition, Figs.~\ref{Fig1:DensPhase}(a) and (b) show the density and phase, respectively, of a pair of at-rest solitons with opposite polarization.
Since each soliton has zero velocity, its core exhibits a $\pi$ phase jump in one component, while the other remains phase-flat.
Notice also that magnetic solitons manifest predominantly in the spin density $s(x)=n_1(x)-n_2(x)$, whereas the total density $n(x)=n_1(x)+n_2(x)$ remains nearly constant.
When this pair is evolved in time, Fig.~\ref{Fig1:DensPhase}(c) shows that the constituent solitons attract and pass through one another, forming an oscillating bound state, whereas solitons with like polarization repel each other [Fig.~\ref{Fig1:DensPhase}(d)].
These contrasting behaviors motivate our systematic study of soliton bound states.
We numerically calculate the interaction potential in both cases, elucidating the mechanism for bound state formation.
We further show that the potential flattens anharmonically with increasing separation, consistent with the diverging oscillation period at larger excitation amplitudes.
We also derive the bound state dissociation energy analytically and confirm quantitative agreement with dynamical simulations.
Finally, we investigate collisions involving soliton bound states, revealing dynamics that range from nearly elastic bounces to soliton swapping, analogous to substitution reactions.

\section{Formalism}

\subsection{System and methods} \label{Sec:system}

We consider two coupled BECs trapped in an elongated geometry, where both components experience the same harmonic potential along the $y$ and $z$ directions, with isotropic trapping frequencies $\omega_y=\omega_z\equiv\omega$, and are unconfined along $x$.
The wavefunction of component $i$ is described by the separable ansatz,
\begin{equation}
\Psi_i(\mathbf{x}) = \psi_i(x)\frac{\exp[-(y^2+z^2)/2l^2]}{l\sqrt{\pi}} \, , \label{Eq:psi}
\end{equation}
with characteristic length $l=\sqrt{\hbar/m\omega}$, where the mass $m$ is assumed to be the same for both components.
Ansatz (\ref{Eq:psi}) is substituted into the two-component Gross-Pitaevskii equation (GPE) \cite{ho1996binary}. The radial part is then integrated out, and subtracting the radial zero-point energy yields the quasi-one-dimensional (quasi-1D) GPEs:
\begin{eqnarray}
   i\hbar\partial_t\psi_1 &=&\left(-\frac{\hbar^2}{2m}\partial_x^2+\frac{g_{11} n_1}{2\pi l^2 }  +\frac{g_{12} n_2 }{2\pi l^2 }    \right)\psi_1\, , \notag\\
   i\hbar\partial_t\psi_2 &=&\left(-\frac{\hbar^2}{2m}\partial_x^2+\frac{g_{12} n_1 }{2\pi l^2 }  +\frac{g_{22} n_2}{2\pi l^2 }    \right)\psi_2\, ,
       \label{Eq:GPE}
\end{eqnarray}
with linear densities $n_j=|\psi_j(x)|^2$ and interaction strengths $g_{ij}=4\pi a_{ij}\hbar^2/m$, where $a_{ij}$ is the $s$-wave scattering length between components $i$ and $j$, and $g_{12}=g_{21}$.

Equation (\ref{Eq:GPE}) is solved with a 4th-order Runge-Kutta method, and stationary solutions are found using imaginary time evolution.
To create zero-velocity solitons, we enforce the phase of the order parameter to exhibit a $\pi$ step function for the appropriate component at the desired position for every imaginary time step, similar to the approach used in Refs.~\cite{pawlowski2015dipolar,bland2017interaction} to create solitons in one-component systems.
We validate this method by evolving the converged solutions in real time. For both isolated at-rest solitons and bound-state pairs produced at stationary points of their inter-soliton potential, we observe no change during the evolution, confirming their stationary state character. 
This method is also useful for creating zero-velocity soliton pairs that are not necessarily stationary states. For example, the initial state ($t=0$) in Fig.~\ref{Fig1:DensPhase}(a) and \ref{Fig1:DensPhase}(a) was created in this way.

\subsection{Parameters and considerations for experiments} \label{Sec:params}

For the purpose of realizing magnetic solitons, we consider the case of a balanced mixture, i.e., $g_{11}=g_{22} \equiv g$, $N_1=N_2\equiv N/2$ and average 1D density per component $n_0 = N/2L$.
We choose a system size $L=30\,\upmu$m and spin-healing length $\xi_s=0.4703\,\upmu$m, where $\xi_s=\hbar/\sqrt{2mn_0\delta g/\pi l^2}$, with $\delta g=g-g_{12}$. 

While our results are given in dimensionless units, they can easily be translated into SI units.
For example, using the mass of $^{87}$Rb, our computational parameters can correspond to $n_0=50\,\upmu \rm{m}^{-1}$, $\omega=2\pi\times4969\,$Hz, and $(a_{11},\,a_{12},\,a_{22}) = (100,95,100)\,a_0$, where $a_0$ is the Bohr radius.
Alternatively, with the choice of $^{23}$Na, our computational parameters can correspond to $n_0=200\,\upmu \rm{m}^{-1}$, $\omega=2\pi\times8611\,$Hz, and
$(a_{11},\,a_{12},\,a_{22}) = (54.54,51.81,54.54)\,a_0$.
These interaction combinations are close to the regimes already realized in magnetic soliton experiments, e.g., see \cite{hoefer2011dark,farolifi2020observation}, and we have checked that our results are qualitatively representative of a broad range of density and trapping strength combinations.

\section{Formulation of the bound state dissociation criterion}\label{Sec:Constant_n}

The order parameter of component $i$  can be parametrized as 
$\psi_i(x)=\sqrt{n_i(x)}e^{i\phi_i(x)}$, where $\phi_i$ is the phase.
For spin excitations and dynamics in two-component BECs with $\delta g \ll g$, one can safely assume that the total density $n$ is uniform (uniform-$n$ approximation) and the dynamics 
are solely characterized by the spin density $s(x)$ and relative phase $\phi_A(x)=\phi_1(x)-\phi_2(x)$ \citep{qu2016magnetic}.
Introducing the polarization distribution, $\cos\theta(x)=s(x)/n$, the GPEs can be reformulated into the
following coupled differential equations \cite{qu2017magnetic}
\begin{eqnarray}
    \partial_\tau\phi_A &=& \cos{\theta}(\partial_\zeta\phi_A)^2-\frac{1}{\sin{\theta}}\partial^2_\zeta\theta-\cos{\theta},
    \notag \\
    \partial_\tau\theta &=& 2\cos{\theta}\,\partial_\zeta\theta\,\partial_\zeta\phi_A+\sin{\theta}\,\partial_\zeta^2\phi_A,
    \label{Eq:CDGL}
\end{eqnarray}
where $\zeta=x/\xi_s$, $\tau=c_s t/\xi_s$ are the dimensionless coordinate and time, respectively, with $c_s=\sqrt{n_0\delta g/2m \pi l^2}$
the speed of spin sound.
This formalism allows the energy of an excitation in the spin density to be evaluated with \cite{qu2017magnetic}
\begin{equation}
E_{\rm{SE}}=\frac{n\hbar c_s}{4}\int d\zeta\left[\left(\frac{\partial\theta}{\partial\zeta}\right)^2+\sin^2\theta\left(\frac{\partial\phi_A}{\partial\zeta}\right)^2+\cos^2\theta\right].
\label{EMS}
\end{equation}

Prior work \cite{qu2016magnetic} revealed a traveling-wave solution, the so-called magnetic soliton, of the form $\theta(\zeta, \tau)\equiv\theta(\zeta - U\tau)$
and $\phi_A(\zeta, \tau)\equiv\phi_A(\zeta- U\tau)$ where $U=v/c_s$ is the speed of the soliton.
The static magnetic soliton $(U=0)$ takes the following simple form
\begin{align}
	 \cos{\theta}&=\pm\frac{1}{\cosh{(\zeta)}} \,, \notag \\
    \phi_A&=\pm\pi\Theta(\zeta)\,,
\end{align}
where $\Theta(\zeta)$ is the Heaviside step function.
We define the $(+)$ solution as spin-up magnetic soliton, since $s(x)>0$ around the core [see soliton on the right in Fig.~\ref{Fig1:DensPhase}(a)].
Conversely, the $(-)$ solution is referred to as a spin-down soliton.
Note that, for $U=0$, the antidark component ($n_i(x)>n_0$ around the core) has no phase variation and the dark component ($n_i(x)<n_0$) exhibits a $\pi$ phase jump.

\begin{figure}
	\centering
	\includegraphics[width=0.38\textwidth]{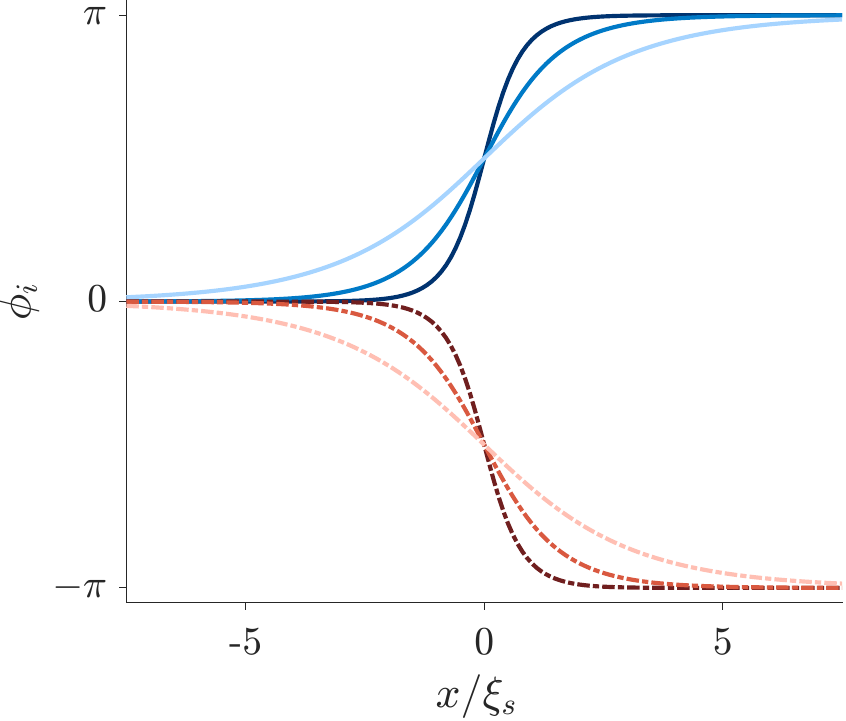}
	\caption{Relative-phase domain walls showing the phase profiles of component 1 (2) as solid (dot-dashed) curves for three different widths, $b=\{0.5, 1, 2\}$, from dark to light.}
	\label{Fig:RPDW}
\end{figure}

An intriguing fact is that a pair of oppositely polarized magnetic solitons, initially cospatial, can be generated with opposite velocities by imprinting a
{\it relative-phase domain wall} of the form \cite{son2002domain}:
\begin{align}
	\cos{\theta(\zeta, \tau=0)}&=0\,, \notag \\
    \phi_A(\zeta, \tau=0)&=4\arctan{e^{\zeta/b}}\,, \label{Eq:RPDW}
\end{align}
where $b$ is its characteristic width. 
Although the spin density is spatially uniform, $\phi_A$ can be symmetrically distributed between the two components, revealing the soliton pairs shown in Fig.~\ref{Fig:RPDW}, where three different widths are considered.
The $\tau>0$ dynamics of the pair falls into one of two categories, as can be predicted by the following energy consideration. 
Using Eq.~(\ref{EMS}), we can obtain the energy of a single magnetic soliton
\begin{equation}
E_\text{MS}^\text{uni}(U)=n\hbar c_s\sqrt{1-U^2}, \label{Eq:MS_Ene}
\end{equation}
and the energy of the relative-phase domain wall
\begin{equation}
	E_\text{RPDW}^\text{uni}(b)=2n\hbar c_s/b.
	\label{Eq:RPDWenergy}
\end{equation}
On the one hand, for large widths $b>1$ (small energies), 
the energy of the initially imprinted relative-phase domain wall can be equal to the sum of the energies of two widely separated solitons moving in opposite directions with $U$ given by
\begin{equation}
U=\pm \sqrt{1-\frac{1}{b^2}}\,.
\end{equation}
On the other hand, when the initial width is narrow, $b<1$, the energy of the generated
soliton pair is too large to form well-separated solitons. Instead, they quickly evolve into a pair of static magnetic solitons ($U=0$) at finite separation, before again moving towards one another, thus forming an oscillatory bound state (analogous to an excited molecule). This counterintuitive behavior may be understood by noting that the energy of a soliton decreases with increasing velocity [see Eq.~(\ref{Eq:MS_Ene})], which corresponds to a negative effective mass 
\footnote{ For comparison, it is interesting to note that while dark solitons also have negative effective masses \cite{pitaevskii2016bose}, certain dark-bright solitons have been predicted to dynamically oscillate between negative and positive effective mass \cite{zhao2020spin}. }
\begin{equation}
m_\text{eff}=\frac{1}{Uc_s^2}\frac{dE_\text{MS}^\text{uni}}{dU}=-\frac{n\hbar}{c_s}\frac{1}{\sqrt{1-U^2}}. \label{Eq:MS_Mass}
\end{equation} 
The boundary between the two regimes is set by the critical width $b_\text{c}=1$, which corresponds to the critical bound-state dissociation energy $E_\text{c}=2n\hbar c_s$.

\section{Numerical results}

\subsection{Soliton-soliton interaction potential}\label{Sec:SSIntPpot}

\begin{figure}[h!]
	\centering
	\includegraphics[width=0.48\textwidth]{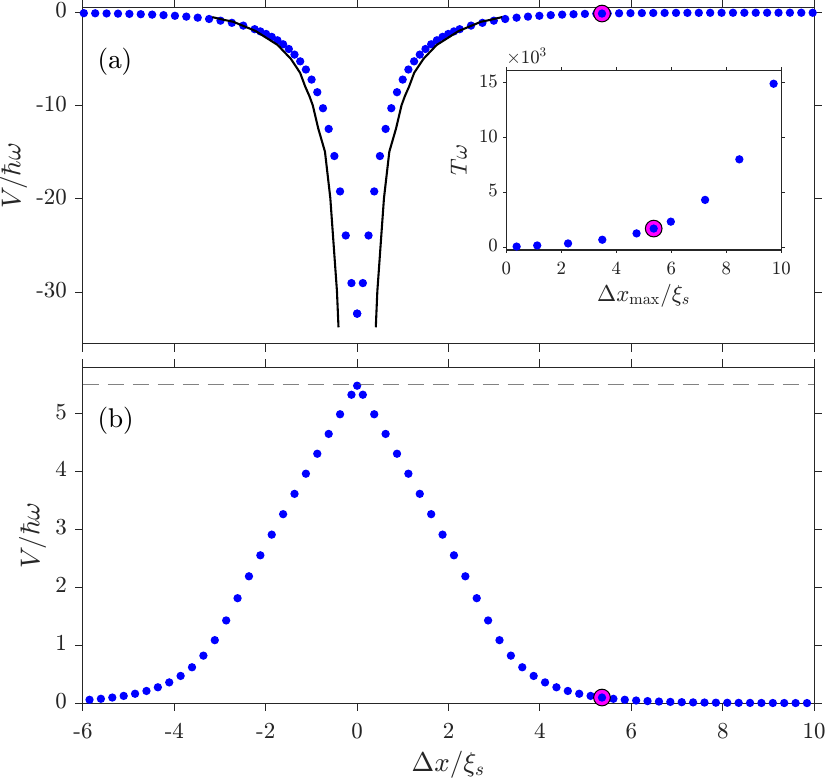}
	\caption{(a) Interaction potential between two magnetic solitons with opposite polarization as a function of their separation, calculated using the GPEs (blue points) and the uniform-$n$ approximation $V^\text{uni}$ (black curve).
Inset shows the oscillation period of the bound states as a function of their maximum separation.
(b) Interaction potential for pairs with like polarization.
When $\Delta x \to 0$, the potential approaches the energy of a single magnetic soliton (horizontal dashed line).
Magenta circles highlight the two cases shown in Fig.~\ref{Fig1:DensPhase}(c,d).
}
	\label{Fig:PotAndT}
\end{figure}

To shed light on the dynamics of magnetic soliton pairs, we define the intersoliton potential as
\begin{equation}
V =\text{sign}(m_\text{eff}) \left[E_\text{2MS}(\Delta x)-E_0\right],
\label{Eq:ISP}
\end{equation}
where $E_\text{2MS}$ is the total system energy for two solitons at rest separated by $\Delta x=x_2-x_1$, with $x_i$ denoting the position of soliton $i$.
The reference energy $E_0$ corresponds to the total system energy for maximally separated solitons at rest, so that $V\to0$ when the solitons are far apart.
The sign prefactor in Eq.~(\ref{Eq:ISP}) is included to make the relationship between the force and acceleration more intuitive.
Since magnetic solitons have a negative effective mass, $\text{sign}(m_\text{eff})=-1$ [see Eq.~(\ref{Eq:MS_Mass})], the acceleration $a =F/m_\text{eff}$ is actually opposite in direction to the force $F$.
However, this sign difference is absorbed into our definition of $V$ and bound states thus occur within potential minima.

Figure \ref{Fig:PotAndT}(a) shows $V$ for two solitons with opposite polarizations. The blue points are calculated using the GPEs following the approach outlined in Sec.~\ref{Sec:system}.
Importantly, there is a minimum at $\Delta x=0$, indicating that the solitons experience an attraction toward zero separation.
This behavior is qualitatively different from that of real molecular potentials, where the minima occur at finite separation.
We have confirmed that zero-separation bound states remain stationary under real time evolution of the GPEs (not shown).
In contrast, Fig.~\ref{Fig:PotAndT}(b) shows $V$ for a pair of solitons with like polarization. 
In the limit $\Delta x\to 0$, $V$ approaches the energy of a single magnetic soliton in the system (horizontal dashed line; see Appendix \ref{Sec:1SolE} for details).
To understand this, recall that the wavefunction phase has two $\pi$ jumps in the same component.
Taking the limit $\Delta x\to 0$ but excluding $\Delta x = 0$, the region where the density is forced to zero approaches a single point, just as for a single soliton.
For increasing $|\Delta x|$, $V$ decreases monotonically, indicating that like-sign solitons repel each other and thus cannot form bound states [see Fig.~\ref{Fig1:DensPhase}(d)].

A semianalytic prediction for the intersoliton potential can be made based on the uniform-$n$ approximation outlined in Sec.~\ref{Sec:Constant_n}.
The soliton interaction energy is obtained from the difference between the energy of a relative-phase domain wall of width $b<1$ and that of two isolated magnetic solitons at zero velocity, i.e.,
\begin{equation}
E^\text{uni} = E_\text{RPDW}^\text{uni}(b)-2E_\text{MS}^\text{uni}(U=0)=2n\hbar c_s\left(\frac{1}{b}-1\right).
\label{Eq:SAE}
\end{equation}
The energy of the relative phase domain wall is equivalent to that of two solitons at rest separated by $\Delta x$, provided that $b<1$, because these two configurations correspond to different phases of the same oscillation.
To determine the relation between $b$ and $\Delta x$, we evolve relative phase domain walls under the uniform-$n$ equations (\ref{Eq:CDGL}) and extract $\Delta x$ as the maximum soliton separation at rest after a quarter of an oscillation period. 
The resulting semianalytic prediction of the potential $V^\text{uni} = \text{sign}(m_\text{eff}) E^\text{uni}$ as a function of $\Delta x$ is plotted in Fig.~\ref{Fig:PotAndT}(a) as solid lines, demonstrating good agreement with the GPE results for moderate soliton separations.
Note, however, that $V^\text{uni}$ diverges for $\Delta x \ll \xi_s$, because the uniform-$n$ constraint significantly overestimates the kinetic energy when $b\ll1$.

The oscillation period $T$ of excited bound states, shown in the inset to Fig.~\ref{Fig:PotAndT}(a), can serve as a useful experimental probe.
We calculate $T$ by evolving the at-rest solitons from the main plot of Fig.~\ref{Fig:PotAndT}(a) as initial states in time-dependent GPE simulations.
The period exhibits anharmonicity, diverging as the oscillation amplitude $\Delta x_\text{max}$ increases and the bound state dissociation energy is approached.
The large-amplitude case highlighted by the magenta circles corresponds to the dynamics in Fig.~\ref{Fig1:DensPhase}(c). For reference, taking the mass of $^{87}$Rb (see Sec.~\ref{Sec:params}) yields a period of $T \approx 60.6\,$ms for this case.

\subsection{Critical binding energy}

\label{CritocalBindingEnergy}

\begin{figure}[h!]
	\centering
	\includegraphics[width=0.48\textwidth]{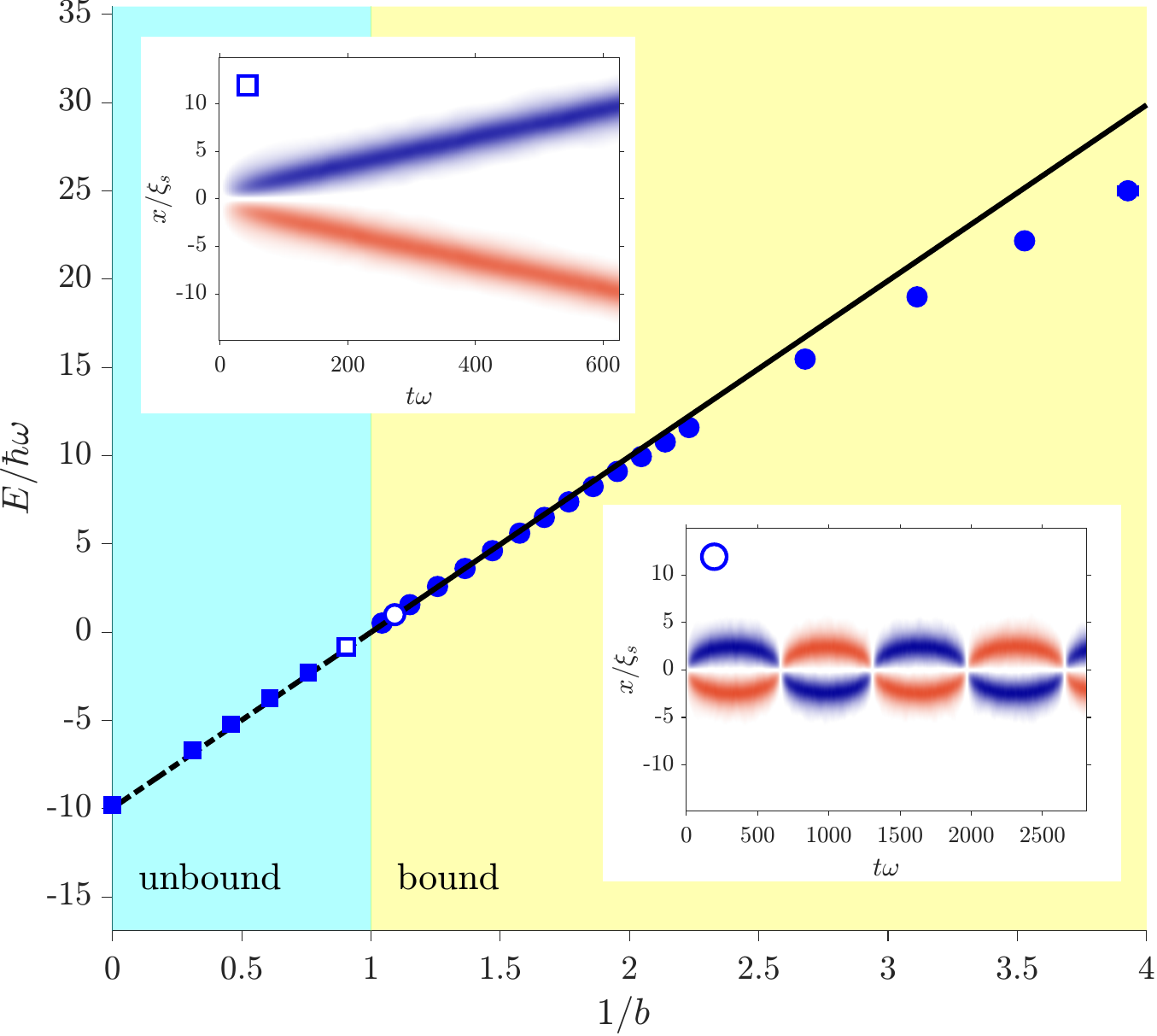} 
	\caption{Energy of relative-phase domain walls---interpreted as cospatial pairs of oppositely polarized magnetic solitons with opposite velocities---as a function of inverse width.
		Symbols show GPE results (circles: bound, squares: unbound), and curves show the uniform-$n$ prediction [Eq.~(\ref{Eq:SAE})] (solid: bound, dashed: unbound).
	Magnetic soliton pairs are bounded (unbounded) when the width of the relative-phase domain wall is smaller (larger) than a critical value, $b_\text{c}=1$ (i.e., $E=0$).
Insets show examples of unbound (left) and bound (right) dynamics starting from the initial states indicated by the open symbols.  } 
	\label{Fig:BindE}
\end{figure}

A relative-phase domain wall is equivalent to a pair of counterpropagating, oppositely polarized solitons.
The soliton speed increases as $1/b$ decreases, while the energy decreases.
In Fig.~\ref{Fig:BindE}, we exploit this by performing dynamical GPE simulations to determine the critical energy below which a soliton pair becomes unbound.

Relative-phase domain walls of width $b_0$ [Eq.~(\ref{Eq:RPDW})] are first imprinted on otherwise uniform BECs.
These are then subjected to a short imaginary-time evolution ($\Delta t=312i/\omega$) to relax the density near the soliton cores, which need not be uniform.
The resulting states, which serve as our initial conditions for real-time evolution, differ slightly in width from $b_0$; we extract the actual $b$ by fitting their phase profiles [Eq.~(\ref{Eq:RPDW})].
The fit errors are small, with error bars in Fig.~\ref{Fig:BindE} generally smaller than the symbols (see Appendix \ref{Sec:fitting}).
The system energy $E_{\rm RPDW}$ is calculated for each state, with the reference energy of two maximally separated solitons subtracted to obtain
\begin{equation}
E = E_{\rm RPDW}(b)-E_0.
\label{Eq:E_GPE_RPDW}
\end{equation}
The states are then evolved in real time to determine whether they remain bound or become unbound.

Figure \ref{Fig:BindE} shows the energy of relative phase domain walls [Eq.~(\ref{Eq:E_GPE_RPDW})] as a function of inverse width.
Squares (circles) denote soliton pairs that evolve to be unbound (bound) in real-time simulations.
Two example simulations are plotted as insets, one on each side of the binding threshold, i.e., $b^{-1}=0.9$ and $b^{-1}=1.1$.
The analytic uniform-$n$ prediction [Eq.~(\ref{Eq:SAE})] agrees quantitatively with the GPE simulations [Eq.~(\ref{Eq:E_GPE_RPDW})] near the threshold.
At larger $1/b$ (slow-moving solitons), deviations arise from the constraint of uniform total density (cf.~Sec.~\ref{Sec:SSIntPpot}).

\subsection{Collisions involving bound states}

\begin{figure}[h!]
	\centering
	\includegraphics[width=0.48\textwidth]{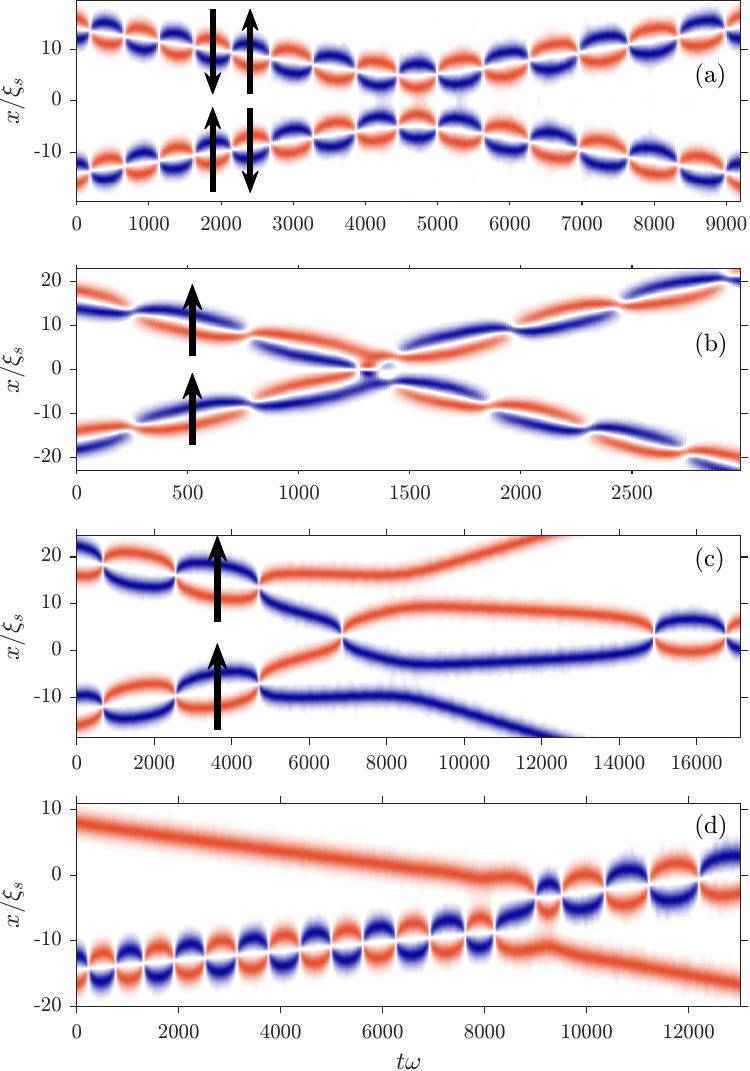}
	\caption{Collisions involving excited bound states under various conditions:
	(a) two tightly bound states with oscillating polarity that remain opposite (black arrows indicate polarity);
	(b) two tightly bound states with the same polarity;
	(c) two loosely bound states with the same polarity;
	(d) an isolated magnetic soliton colliding with a bound state.
	} 
	\label{Fig:BounStateCol}
\end{figure}

A well-known property of solitons is their ability to survive collisions unscathed.
Because this work does not focus on the Manakov limit, which requires $\Delta g = 0$, our system is not perfectly integrable, and we therefore do not expect perfectly elastic collisions.
Nevertheless, it is of interest to test whether magnetic soliton bound states can remain intact after collisions.
In realistic experimental settings, isolated bound states are likely to occur in excited configurations. Accordingly, in Fig.~\ref{Fig:BounStateCol} we investigate collisions involving excited bound states under various conditions.

Since solitons of like polarization repel and those of opposite polarization attract, an excited bound state behaves like a dipole oscillating with period $T$.
The time-dependent polarity of these dipoles is illustrated in Fig.~\ref{Fig:BounStateCol} by black arrows pointing from spin-down (red) to spin-up (blue) solitons within each bound state.
Figure \ref{Fig:BounStateCol}(a) shows a collision between two tightly bound states, each with an initial $T = 31.04\,$ms.
In this case, the polarities of the two bound states remain opposite throughout the collision, leading to inter bound-state repulsion and a clear bounce, with both bound states remaining intact.
Evidence for bound state energy loss is seen in the slightly increasing oscillation periods over time.
 The energy lost is converted to phonons, which are not visible with this color scheme.

Figure \ref{Fig:BounStateCol}(b) shows two tightly bound states (initially $T=34.2\,$ms) with the same polarity (black arrows pointing in the same direction).
In this case, the bound-state dipoles tend to attract each other, yet both bound states nevertheless survive the collision.
This attractive interaction is only weakly visible due to the tight binding and the relatively fast center-of-mass motions of the bound states.
By contrast, the attractive interaction between same-polarity bound states has striking consequences for two loosely bound states (initially $T=128.9\,$ms), as shown in Fig.~\ref{Fig:BounStateCol}(c).
Here, the inner solitons---having opposite polarization---not only cross each other but are actually torn from their respective bound states and form a new bound state, while the two outer solitons become unbound.
Finally, in the case of a single soliton colliding with a bound state (initially $T = 36.59\,$ms), Fig.~\ref{Fig:BounStateCol}(d) illustrates a type of substitution reaction, in which the newcomer replaces one of the bound state's solitons and ejects it in the process.

\section{Discussion}

In this paper, we investigated the mechanisms underlying the formation of magnetic soliton bound states and their collision dynamics.
By calculating the inter-soliton potential, we showed that magnetic solitons with opposite polarization can form bound molecular states, whereas those with like polarization cannot.
The analytically determined dissociation energy of such bound states shows excellent agreement with numerical simulations.
The dynamics of bound states exhibit a diverging oscillation period as the unbinding threshold is approached.
Collisions between excited bound states exhibit dipole-like interactions, leading to attraction or repulsion depending on the relative dipole polarity.
We also demonstrated substitution-type reactions, in which one soliton of a bound state is replaced by another that collides with it.

Future studies could investigate bound states in binary dipolar BECs, where long-range interactions and the presence of a roton-maxon spectrum may qualitatively affect the inter-soliton potentials and dynamics.
Another interesting direction is the generalization to Rabi-coupled systems, which exhibit rich magnetic-soliton dynamics. 
Furthermore, one could study the crossover from the quasi-1D to the three-dimensional cigar-shaped regime.
While magnetic solitons are expected to be more robust against three-dimensional instabilities than their single-component counterparts---since their size is governed by the spin-healing length, which grows large near the immiscibility threshold---it would be valuable to investigate the two-component analog of snake instabilities and the formation of vortex rings.
More generally, dark-antidark solitons also occur in imbalanced binary BECs, where the intraspecies interactions of the components differ \cite{congy2016dispersive}. We have confirmed that bound states also arise in such imbalanced systems (up to at least $g_{11}/g_{22}=1.2$; not shown); a systematic study of these bound states represents another interesting research direction.
From an experimental perspective, oscillations of bound states, as well as collisions involving magnetic solitons and their molecular counterparts, could provide direct information about the interaction potentials, offering a route to test the predictions of this work.

\section*{Acknowledgements}
We thank T. Billam, T. Bland, F. Ferlaino, L. Bellinato Giacomelli, A. Madhusudan, N. Masalaeva, N. Parker, E. Poli, and P. Senarath Yapa for fruitful discussions.
This research was funded in part by the Austrian Science Fund (FWF) [Grant DOI: 10.55776/P 36850-N]. C.~Q.~is supported by ACC New Jersey under Contract No. W15QKN-18-D-0040.

\appendix

\section{Numerical calculation of single-soliton energy} 
\label{Sec:1SolE}

Our numerical GPE simulations do not assume a constant total density $n$ and are performed in a finite-sized system. For these reasons, rather than using Eq.~(\ref{Eq:MS_Ene}), we numerically calculate the energy of a single soliton in our system.
Due to the periodic boundary conditions, our imprinting protocol does not permit the creation of a single isolated soliton.
Instead, we consider a system of length $2L$ containing two maximally separated solitons (with energy $E_{0}^{2L}$), which has the same number of solitons per unit length as a system of size $L$ with only one soliton.
The numerically calculated energy associated with a single soliton is then defined as
\begin{equation}
E_\text{MS}=   \text{sign}(m_\text{eff}) \left( E_{0}^{2L}/2 - E_{0} \right) , \label{Eq:E_MS_GPE}
\end{equation}
where $E_0$ is the same reference energy used in Eq.~(\ref{Eq:ISP}).
Equation (\ref{Eq:E_MS_GPE}) is plotted as the horizontal dashed line in Fig.~\ref{Fig:PotAndT}(b). Note that the difference between Eq.~(\ref{Eq:E_MS_GPE}) and Eq.~(\ref{Eq:MS_Ene}) is approximately 7$\%$ for $U=0$.

Alternatively, recall that for like-sign solitons in the limit $\Delta x \to 0$, the inter-soliton potential $V$ [Eq.~(\ref{Eq:ISP})] approaches the energy of a single magnetic soliton.
In Fig.\ref{Fig:PotAndT}(b), a single data point represents $V(\Delta x \to 0)$.
This value cannot be obtained using realistic numerical grids.
Instead, we employ a numerical workaround: during each step of imaginary-time evolution in a system of length $L$, the density is artificially constrained to zero for only a single grid point.
This procedure yields excellent agreement with the single-soliton energy $E_\text{MS}$.

\begin{figure}[h!]
	\centering
	\includegraphics[width=0.47\textwidth]{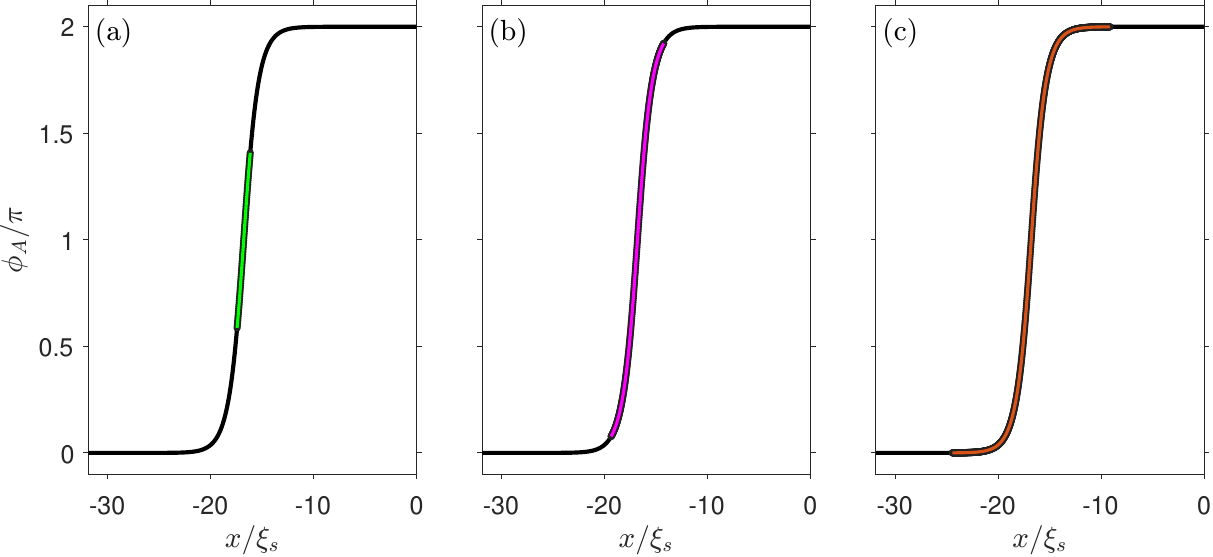}
	\caption{Example of the phase-fitting procedure used to obtain the parameter $b$ for the initial states described in Sec.~\ref{CritocalBindingEnergy}. 
Fits were performed with widths: (a) $x_b-x_a$, (b) $4(x_b-x_a)$, and (c) $12(x_b-x_a)$, where $x_b-x_a = 1.511\,\xi_s$. The resulting fitted values are $b_A = 1.0896$, $b_B = 1.0926$, and $b_C = 1.0930$, respectively.
	 }
	\label{Fig:Phasefitting}
\end{figure}

\section{Fitting of relative phase domain walls to find their widths} \label{Sec:fitting}

In Sec.~\ref{CritocalBindingEnergy}, we extract the relative phase domain wall width $ b $ by fitting the relative phase $ \phi_A $ of the initial states to Eq.~(\ref{Eq:RPDW}). The fit is performed using three different fitting widths: $ x_b - x_a $, $ 4(x_b - x_a) $, and $ 12(x_b - x_a) $, with $x_a$ and $x_b$ being the positions where $\phi_A(x_a) = 0.6\pi$ and $\phi_A(x_b) = 1.4\pi$, respectively (see Fig.~\ref{Fig:Phasefitting}). The final value of $ b $ is taken as the mean of the three fits, and the error is given by the standard deviation. An example is shown in Fig.~\ref{Fig:Phasefitting}.


\begin{thebibliography}{40}%
\makeatletter
\providecommand \@ifxundefined [1]{%
 \@ifx{#1\undefined}
}%
\providecommand \@ifnum [1]{%
 \ifnum #1\expandafter \@firstoftwo
 \else \expandafter \@secondoftwo
 \fi
}%
\providecommand \@ifx [1]{%
 \ifx #1\expandafter \@firstoftwo
 \else \expandafter \@secondoftwo
 \fi
}%
\providecommand \natexlab [1]{#1}%
\providecommand \enquote  [1]{``#1''}%
\providecommand \bibnamefont  [1]{#1}%
\providecommand \bibfnamefont [1]{#1}%
\providecommand \citenamefont [1]{#1}%
\providecommand \href@noop [0]{\@secondoftwo}%
\providecommand \href [0]{\begingroup \@sanitize@url \@href}%
\providecommand \@href[1]{\@@startlink{#1}\@@href}%
\providecommand \@@href[1]{\endgroup#1\@@endlink}%
\providecommand \@sanitize@url [0]{\catcode `\\12\catcode `\$12\catcode
  `\&12\catcode `\#12\catcode `\^12\catcode `\_12\catcode `\%12\relax}%
\providecommand \@@startlink[1]{}%
\providecommand \@@endlink[0]{}%
\providecommand \url  [0]{\begingroup\@sanitize@url \@url }%
\providecommand \@url [1]{\endgroup\@href {#1}{\urlprefix }}%
\providecommand \urlprefix  [0]{URL }%
\providecommand \Eprint [0]{\href }%
\providecommand \doibase [0]{https://doi.org/}%
\providecommand \selectlanguage [0]{\@gobble}%
\providecommand \bibinfo  [0]{\@secondoftwo}%
\providecommand \bibfield  [0]{\@secondoftwo}%
\providecommand \translation [1]{[#1]}%
\providecommand \BibitemOpen [0]{}%
\providecommand \bibitemStop [0]{}%
\providecommand \bibitemNoStop [0]{.\EOS\space}%
\providecommand \EOS [0]{\spacefactor3000\relax}%
\providecommand \BibitemShut  [1]{\csname bibitem#1\endcsname}%
\let\auto@bib@innerbib\@empty
\bibitem [{\citenamefont {Russell}(1885)}]{Russell1885the}%
  \BibitemOpen
  \bibfield  {author} {\bibinfo {author} {\bibfnamefont {J.~S.}\ \bibnamefont
  {Russell}},\ }\href@noop {} {\emph {\bibinfo {title} {The Wave of Translation
  in the Oceans of Water}}}\ (\bibinfo  {publisher} {Tr\"ubner \& Company},\
  \bibinfo {address} {London},\ \bibinfo {year} {1885})\BibitemShut {NoStop}%
\bibitem [{\citenamefont {Ablowitz}(2011)}]{Ablowitz2011nonlinear}%
  \BibitemOpen
  \bibfield  {author} {\bibinfo {author} {\bibfnamefont {M.}~\bibnamefont
  {Ablowitz}},\ }\href@noop {} {\emph {\bibinfo {title} {Nonlinear Dispersive
  Waves, Asymptotic Analysis and Solitons}}}\ (\bibinfo  {publisher} {Cambridge
  University Press},\ \bibinfo {address} {New York},\ \bibinfo {year}
  {2011})\BibitemShut {NoStop}%
\bibitem [{\citenamefont {Kivshar}\ and\ \citenamefont
  {Agrawal}(2003)}]{Kivshar2003optical}%
  \BibitemOpen
  \bibfield  {author} {\bibinfo {author} {\bibfnamefont {Y.~S.}\ \bibnamefont
  {Kivshar}}\ and\ \bibinfo {author} {\bibfnamefont {G.~P.}\ \bibnamefont
  {Agrawal}},\ }\href@noop {} {\emph {\bibinfo {title} {Optical Solitons: From
  Fibers to Photonic Crystal}}}\ (\bibinfo  {publisher} {Academic Press},\
  \bibinfo {address} {San Diego},\ \bibinfo {year} {2003})\BibitemShut
  {NoStop}%
\bibitem [{\citenamefont {Kono}\ and\ \citenamefont
  {Skori\'cc}(2010)}]{Kono2010nonlinear}%
  \BibitemOpen
  \bibfield  {author} {\bibinfo {author} {\bibfnamefont {M.}~\bibnamefont
  {Kono}}\ and\ \bibinfo {author} {\bibfnamefont {M.}~\bibnamefont
  {Skori\'cc}},\ }\href@noop {} {\emph {\bibinfo {title} {Nonlinear Physics of
  Plasmas}}}\ (\bibinfo  {publisher} {Springer},\ \bibinfo {address} {Berlin,
  Heidelberg},\ \bibinfo {year} {2010})\BibitemShut {NoStop}%
\bibitem [{\citenamefont {Heeger}\ \emph {et~al.}(1988)\citenamefont {Heeger},
  \citenamefont {Kivelson}, \citenamefont {Schrieffer},\ and\ \citenamefont
  {Su}}]{Heeger1988solitons}%
  \BibitemOpen
  \bibfield  {author} {\bibinfo {author} {\bibfnamefont {A.~J.}\ \bibnamefont
  {Heeger}}, \bibinfo {author} {\bibfnamefont {S.}~\bibnamefont {Kivelson}},
  \bibinfo {author} {\bibfnamefont {J.~R.}\ \bibnamefont {Schrieffer}},\ and\
  \bibinfo {author} {\bibfnamefont {W.~P.}\ \bibnamefont {Su}},\ }\bibfield
  {title} {\bibinfo {title} {Solitons in conducting polymers},\ }\href
  {https://doi.org/10.1103/RevModPhys.60.781} {\bibfield  {journal} {\bibinfo
  {journal} {Rev. Mod. Phys.}\ }\textbf {\bibinfo {volume} {60}},\ \bibinfo
  {pages} {781} (\bibinfo {year} {1988})}\BibitemShut {NoStop}%
\bibitem [{\citenamefont {Dauxois}\ and\ \citenamefont
  {Peyrard}(2006)}]{Dauxois2006physics}%
  \BibitemOpen
  \bibfield  {author} {\bibinfo {author} {\bibfnamefont {T.}~\bibnamefont
  {Dauxois}}\ and\ \bibinfo {author} {\bibfnamefont {M.}~\bibnamefont
  {Peyrard}},\ }\href@noop {} {\emph {\bibinfo {title} {Physics of Solitons}}}\
  (\bibinfo  {publisher} {Cambridge University Press},\ \bibinfo {address}
  {Cambridge},\ \bibinfo {year} {2006})\BibitemShut {NoStop}%
\bibitem [{\citenamefont {Burger}\ \emph {et~al.}(1999)\citenamefont {Burger},
  \citenamefont {Bongs}, \citenamefont {Dettmer}, \citenamefont {Ertmer},
  \citenamefont {Sengstock}, \citenamefont {Sanpera}, \citenamefont
  {Shlyapnikov},\ and\ \citenamefont {Lewenstein}}]{Burger1999dark}%
  \BibitemOpen
  \bibfield  {author} {\bibinfo {author} {\bibfnamefont {S.}~\bibnamefont
  {Burger}}, \bibinfo {author} {\bibfnamefont {K.}~\bibnamefont {Bongs}},
  \bibinfo {author} {\bibfnamefont {S.}~\bibnamefont {Dettmer}}, \bibinfo
  {author} {\bibfnamefont {W.}~\bibnamefont {Ertmer}}, \bibinfo {author}
  {\bibfnamefont {K.}~\bibnamefont {Sengstock}}, \bibinfo {author}
  {\bibfnamefont {A.}~\bibnamefont {Sanpera}}, \bibinfo {author} {\bibfnamefont
  {G.~V.}\ \bibnamefont {Shlyapnikov}},\ and\ \bibinfo {author} {\bibfnamefont
  {M.}~\bibnamefont {Lewenstein}},\ }\bibfield  {title} {\bibinfo {title} {Dark
  solitons in bose-einstein condensates},\ }\href
  {https://doi.org/10.1103/PhysRevLett.83.5198} {\bibfield  {journal} {\bibinfo
   {journal} {Phys. Rev. Lett.}\ }\textbf {\bibinfo {volume} {83}},\ \bibinfo
  {pages} {5198} (\bibinfo {year} {1999})}\BibitemShut {NoStop}%
\bibitem [{\citenamefont {Denschlag}\ \emph {et~al.}(2000)\citenamefont
  {Denschlag}, \citenamefont {Simsarian}, \citenamefont {Feder}, \citenamefont
  {Clark}, \citenamefont {Collins}, \citenamefont {Cubizolles}, \citenamefont
  {Deng}, \citenamefont {Hagley}, \citenamefont {Helmerson}, \citenamefont
  {Reinhardt} \emph {et~al.}}]{denschlag2000generating}%
  \BibitemOpen
  \bibfield  {author} {\bibinfo {author} {\bibfnamefont {J.}~\bibnamefont
  {Denschlag}}, \bibinfo {author} {\bibfnamefont {J.~E.}\ \bibnamefont
  {Simsarian}}, \bibinfo {author} {\bibfnamefont {D.~L.}\ \bibnamefont
  {Feder}}, \bibinfo {author} {\bibfnamefont {C.~W.}\ \bibnamefont {Clark}},
  \bibinfo {author} {\bibfnamefont {L.~A.}\ \bibnamefont {Collins}}, \bibinfo
  {author} {\bibfnamefont {J.}~\bibnamefont {Cubizolles}}, \bibinfo {author}
  {\bibfnamefont {L.}~\bibnamefont {Deng}}, \bibinfo {author} {\bibfnamefont
  {E.~W.}\ \bibnamefont {Hagley}}, \bibinfo {author} {\bibfnamefont
  {K.}~\bibnamefont {Helmerson}}, \bibinfo {author} {\bibfnamefont {W.~P.}\
  \bibnamefont {Reinhardt}}, \emph {et~al.},\ }\bibfield  {title} {\bibinfo
  {title} {Generating solitons by phase engineering of a bose-einstein
  condensate},\ }\href@noop {} {\bibfield  {journal} {\bibinfo  {journal}
  {Science}\ }\textbf {\bibinfo {volume} {287}},\ \bibinfo {pages} {97}
  (\bibinfo {year} {2000})}\BibitemShut {NoStop}%
\bibitem [{\citenamefont {Dutton}\ \emph {et~al.}(2001)\citenamefont {Dutton},
  \citenamefont {Budde}, \citenamefont {Slowe},\ and\ \citenamefont
  {Hau}}]{dutton2001observation}%
  \BibitemOpen
  \bibfield  {author} {\bibinfo {author} {\bibfnamefont {Z.}~\bibnamefont
  {Dutton}}, \bibinfo {author} {\bibfnamefont {M.}~\bibnamefont {Budde}},
  \bibinfo {author} {\bibfnamefont {C.}~\bibnamefont {Slowe}},\ and\ \bibinfo
  {author} {\bibfnamefont {L.~V.}\ \bibnamefont {Hau}},\ }\bibfield  {title}
  {\bibinfo {title} {Observation of quantum shock waves created with
  ultra-compressed slow light pulses in a bose-einstein condensate},\
  }\href@noop {} {\bibfield  {journal} {\bibinfo  {journal} {Science}\ }\textbf
  {\bibinfo {volume} {293}},\ \bibinfo {pages} {663} (\bibinfo {year}
  {2001})}\BibitemShut {NoStop}%
\bibitem [{\citenamefont {Anderson}\ \emph {et~al.}(2001)\citenamefont
  {Anderson}, \citenamefont {Haljan}, \citenamefont {Regal}, \citenamefont
  {Feder}, \citenamefont {Collins}, \citenamefont {Clark},\ and\ \citenamefont
  {Cornell}}]{Anderson2001watching}%
  \BibitemOpen
  \bibfield  {author} {\bibinfo {author} {\bibfnamefont {B.~P.}\ \bibnamefont
  {Anderson}}, \bibinfo {author} {\bibfnamefont {P.~C.}\ \bibnamefont
  {Haljan}}, \bibinfo {author} {\bibfnamefont {C.~A.}\ \bibnamefont {Regal}},
  \bibinfo {author} {\bibfnamefont {D.~L.}\ \bibnamefont {Feder}}, \bibinfo
  {author} {\bibfnamefont {L.~A.}\ \bibnamefont {Collins}}, \bibinfo {author}
  {\bibfnamefont {C.~W.}\ \bibnamefont {Clark}},\ and\ \bibinfo {author}
  {\bibfnamefont {E.~A.}\ \bibnamefont {Cornell}},\ }\bibfield  {title}
  {\bibinfo {title} {Watching dark solitons decay into vortex rings in a
  bose-einstein condensate},\ }\href
  {https://doi.org/10.1103/PhysRevLett.86.2926} {\bibfield  {journal} {\bibinfo
   {journal} {Phys. Rev. Lett.}\ }\textbf {\bibinfo {volume} {86}},\ \bibinfo
  {pages} {2926} (\bibinfo {year} {2001})}\BibitemShut {NoStop}%
\bibitem [{\citenamefont {Strecker}\ \emph {et~al.}(2002)\citenamefont
  {Strecker}, \citenamefont {Partridge}, \citenamefont {Truscott},\ and\
  \citenamefont {Hulet}}]{strecker2002formation}%
  \BibitemOpen
  \bibfield  {author} {\bibinfo {author} {\bibfnamefont {K.~E.}\ \bibnamefont
  {Strecker}}, \bibinfo {author} {\bibfnamefont {G.~B.}\ \bibnamefont
  {Partridge}}, \bibinfo {author} {\bibfnamefont {A.~G.}\ \bibnamefont
  {Truscott}},\ and\ \bibinfo {author} {\bibfnamefont {R.~G.}\ \bibnamefont
  {Hulet}},\ }\bibfield  {title} {\bibinfo {title} {Formation and propagation
  of matter-wave soliton trains},\ }\href@noop {} {\bibfield  {journal}
  {\bibinfo  {journal} {Nature}\ }\textbf {\bibinfo {volume} {417}},\ \bibinfo
  {pages} {150} (\bibinfo {year} {2002})}\BibitemShut {NoStop}%
\bibitem [{\citenamefont {Khaykovich}\ \emph {et~al.}(2002)\citenamefont
  {Khaykovich}, \citenamefont {Schreck}, \citenamefont {Ferrari}, \citenamefont
  {Bourdel}, \citenamefont {Cubizolles}, \citenamefont {Carr}, \citenamefont
  {Castin},\ and\ \citenamefont {Salomon}}]{khaykovich2002formation}%
  \BibitemOpen
  \bibfield  {author} {\bibinfo {author} {\bibfnamefont {L.}~\bibnamefont
  {Khaykovich}}, \bibinfo {author} {\bibfnamefont {F.}~\bibnamefont {Schreck}},
  \bibinfo {author} {\bibfnamefont {G.}~\bibnamefont {Ferrari}}, \bibinfo
  {author} {\bibfnamefont {T.}~\bibnamefont {Bourdel}}, \bibinfo {author}
  {\bibfnamefont {J.}~\bibnamefont {Cubizolles}}, \bibinfo {author}
  {\bibfnamefont {L.~D.}\ \bibnamefont {Carr}}, \bibinfo {author}
  {\bibfnamefont {Y.}~\bibnamefont {Castin}},\ and\ \bibinfo {author}
  {\bibfnamefont {C.}~\bibnamefont {Salomon}},\ }\bibfield  {title} {\bibinfo
  {title} {Formation of a matter-wave bright soliton},\ }\href@noop {}
  {\bibfield  {journal} {\bibinfo  {journal} {Science}\ }\textbf {\bibinfo
  {volume} {296}},\ \bibinfo {pages} {1290} (\bibinfo {year}
  {2002})}\BibitemShut {NoStop}%
\bibitem [{Note1()}]{Note1}%
  \BibitemOpen
  \bibinfo {note} {It is worth noting that, although vector solitons may not
  always be true solitons in the strict mathematical sense, they possess key
  soliton characteristics, and we will refer to them as solitons for
  consistency with the literature.}\BibitemShut {Stop}%
\bibitem [{\citenamefont {Myatt}\ \emph {et~al.}(1997)\citenamefont {Myatt},
  \citenamefont {Burt}, \citenamefont {Ghrist}, \citenamefont {Cornell},\ and\
  \citenamefont {Wieman}}]{myatt1997production}%
  \BibitemOpen
  \bibfield  {author} {\bibinfo {author} {\bibfnamefont {C.}~\bibnamefont
  {Myatt}}, \bibinfo {author} {\bibfnamefont {E.}~\bibnamefont {Burt}},
  \bibinfo {author} {\bibfnamefont {R.}~\bibnamefont {Ghrist}}, \bibinfo
  {author} {\bibfnamefont {E.~A.}\ \bibnamefont {Cornell}},\ and\ \bibinfo
  {author} {\bibfnamefont {C.}~\bibnamefont {Wieman}},\ }\bibfield  {title}
  {\bibinfo {title} {Production of two overlapping bose-einstein condensates by
  sympathetic cooling},\ }\href@noop {} {\bibfield  {journal} {\bibinfo
  {journal} {Physical Review Letters}\ }\textbf {\bibinfo {volume} {78}},\
  \bibinfo {pages} {586} (\bibinfo {year} {1997})}\BibitemShut {NoStop}%
\bibitem [{\citenamefont {Hall}\ \emph {et~al.}(1998)\citenamefont {Hall},
  \citenamefont {Matthews}, \citenamefont {Ensher}, \citenamefont {Wieman},\
  and\ \citenamefont {Cornell}}]{hall1998dynamics}%
  \BibitemOpen
  \bibfield  {author} {\bibinfo {author} {\bibfnamefont {D.}~\bibnamefont
  {Hall}}, \bibinfo {author} {\bibfnamefont {M.}~\bibnamefont {Matthews}},
  \bibinfo {author} {\bibfnamefont {J.}~\bibnamefont {Ensher}}, \bibinfo
  {author} {\bibfnamefont {C.}~\bibnamefont {Wieman}},\ and\ \bibinfo {author}
  {\bibfnamefont {E.~A.}\ \bibnamefont {Cornell}},\ }\bibfield  {title}
  {\bibinfo {title} {Dynamics of component separation in a binary mixture of
  bose-einstein condensates},\ }\href@noop {} {\bibfield  {journal} {\bibinfo
  {journal} {Physical Review Letters}\ }\textbf {\bibinfo {volume} {81}},\
  \bibinfo {pages} {1539} (\bibinfo {year} {1998})}\BibitemShut {NoStop}%
\bibitem [{\citenamefont {Maddaloni}\ \emph {et~al.}(2000)\citenamefont
  {Maddaloni}, \citenamefont {Modugno}, \citenamefont {Fort}, \citenamefont
  {Minardi},\ and\ \citenamefont {Inguscio}}]{maddaloni2000collective}%
  \BibitemOpen
  \bibfield  {author} {\bibinfo {author} {\bibfnamefont {P.}~\bibnamefont
  {Maddaloni}}, \bibinfo {author} {\bibfnamefont {M.}~\bibnamefont {Modugno}},
  \bibinfo {author} {\bibfnamefont {C.}~\bibnamefont {Fort}}, \bibinfo {author}
  {\bibfnamefont {F.}~\bibnamefont {Minardi}},\ and\ \bibinfo {author}
  {\bibfnamefont {M.}~\bibnamefont {Inguscio}},\ }\bibfield  {title} {\bibinfo
  {title} {Collective oscillations of two colliding bose-einstein
  condensates},\ }\href@noop {} {\bibfield  {journal} {\bibinfo  {journal}
  {Physical review letters}\ }\textbf {\bibinfo {volume} {85}},\ \bibinfo
  {pages} {2413} (\bibinfo {year} {2000})}\BibitemShut {NoStop}%
\bibitem [{\citenamefont {Papp}\ \emph {et~al.}(2008)\citenamefont {Papp},
  \citenamefont {Pino},\ and\ \citenamefont {Wieman}}]{papp2008tunable}%
  \BibitemOpen
  \bibfield  {author} {\bibinfo {author} {\bibfnamefont {S.}~\bibnamefont
  {Papp}}, \bibinfo {author} {\bibfnamefont {J.}~\bibnamefont {Pino}},\ and\
  \bibinfo {author} {\bibfnamefont {C.}~\bibnamefont {Wieman}},\ }\bibfield
  {title} {\bibinfo {title} {Tunable miscibility in a dual-species
  bose-einstein condensate},\ }\href@noop {} {\bibfield  {journal} {\bibinfo
  {journal} {Physical review letters}\ }\textbf {\bibinfo {volume} {101}},\
  \bibinfo {pages} {040402} (\bibinfo {year} {2008})}\BibitemShut {NoStop}%
\bibitem [{\citenamefont {Thalhammer}\ \emph {et~al.}(2008)\citenamefont
  {Thalhammer}, \citenamefont {Barontini}, \citenamefont {De~Sarlo},
  \citenamefont {Catani}, \citenamefont {Minardi},\ and\ \citenamefont
  {Inguscio}}]{thalhammer2008double}%
  \BibitemOpen
  \bibfield  {author} {\bibinfo {author} {\bibfnamefont {G.}~\bibnamefont
  {Thalhammer}}, \bibinfo {author} {\bibfnamefont {G.}~\bibnamefont
  {Barontini}}, \bibinfo {author} {\bibfnamefont {L.}~\bibnamefont {De~Sarlo}},
  \bibinfo {author} {\bibfnamefont {J.}~\bibnamefont {Catani}}, \bibinfo
  {author} {\bibfnamefont {F.}~\bibnamefont {Minardi}},\ and\ \bibinfo {author}
  {\bibfnamefont {M.}~\bibnamefont {Inguscio}},\ }\bibfield  {title} {\bibinfo
  {title} {Double species bose-einstein condensate with tunable interspecies
  interactions},\ }\href@noop {} {\bibfield  {journal} {\bibinfo  {journal}
  {Physical review letters}\ }\textbf {\bibinfo {volume} {100}},\ \bibinfo
  {pages} {210402} (\bibinfo {year} {2008})}\BibitemShut {NoStop}%
\bibitem [{\citenamefont {Becker}\ \emph {et~al.}(2008)\citenamefont {Becker},
  \citenamefont {Stellmer}, \citenamefont {Soltan-Panahi}, \citenamefont
  {D{\"o}rscher}, \citenamefont {Baumert}, \citenamefont {Richter},
  \citenamefont {Kronj{\"a}ger}, \citenamefont {Bongs},\ and\ \citenamefont
  {Sengstock}}]{becker2008oscillations}%
  \BibitemOpen
  \bibfield  {author} {\bibinfo {author} {\bibfnamefont {C.}~\bibnamefont
  {Becker}}, \bibinfo {author} {\bibfnamefont {S.}~\bibnamefont {Stellmer}},
  \bibinfo {author} {\bibfnamefont {P.}~\bibnamefont {Soltan-Panahi}}, \bibinfo
  {author} {\bibfnamefont {S.}~\bibnamefont {D{\"o}rscher}}, \bibinfo {author}
  {\bibfnamefont {M.}~\bibnamefont {Baumert}}, \bibinfo {author} {\bibfnamefont
  {E.-M.}\ \bibnamefont {Richter}}, \bibinfo {author} {\bibfnamefont
  {J.}~\bibnamefont {Kronj{\"a}ger}}, \bibinfo {author} {\bibfnamefont
  {K.}~\bibnamefont {Bongs}},\ and\ \bibinfo {author} {\bibfnamefont
  {K.}~\bibnamefont {Sengstock}},\ }\bibfield  {title} {\bibinfo {title}
  {Oscillations and interactions of dark and dark--bright solitons in
  bose--einstein condensates},\ }\href@noop {} {\bibfield  {journal} {\bibinfo
  {journal} {Nature Physics}\ }\textbf {\bibinfo {volume} {4}},\ \bibinfo
  {pages} {496} (\bibinfo {year} {2008})}\BibitemShut {NoStop}%
\bibitem [{\citenamefont {Hoefer}\ \emph {et~al.}(2011)\citenamefont {Hoefer},
  \citenamefont {Chang}, \citenamefont {Hamner},\ and\ \citenamefont
  {Engels}}]{hoefer2011dark}%
  \BibitemOpen
  \bibfield  {author} {\bibinfo {author} {\bibfnamefont {M.~A.}\ \bibnamefont
  {Hoefer}}, \bibinfo {author} {\bibfnamefont {J.~J.}\ \bibnamefont {Chang}},
  \bibinfo {author} {\bibfnamefont {C.}~\bibnamefont {Hamner}},\ and\ \bibinfo
  {author} {\bibfnamefont {P.}~\bibnamefont {Engels}},\ }\bibfield  {title}
  {\bibinfo {title} {Dark-dark solitons and modulational instability in
  miscible two-component bose-einstein condensates},\ }\href
  {https://doi.org/10.1103/PhysRevA.84.041605} {\bibfield  {journal} {\bibinfo
  {journal} {Phys. Rev. A}\ }\textbf {\bibinfo {volume} {84}},\ \bibinfo
  {pages} {041605} (\bibinfo {year} {2011})}\BibitemShut {NoStop}%
\bibitem [{\citenamefont {Yan}\ \emph {et~al.}(2012)\citenamefont {Yan},
  \citenamefont {Chang}, \citenamefont {Hamner}, \citenamefont {Hoefer},
  \citenamefont {Kevrekidis}, \citenamefont {Engels}, \citenamefont
  {Achilleos}, \citenamefont {Frantzeskakis},\ and\ \citenamefont
  {Cuevas}}]{yan2012beating}%
  \BibitemOpen
  \bibfield  {author} {\bibinfo {author} {\bibfnamefont {D.}~\bibnamefont
  {Yan}}, \bibinfo {author} {\bibfnamefont {J.}~\bibnamefont {Chang}}, \bibinfo
  {author} {\bibfnamefont {C.}~\bibnamefont {Hamner}}, \bibinfo {author}
  {\bibfnamefont {M.}~\bibnamefont {Hoefer}}, \bibinfo {author} {\bibfnamefont
  {P.~G.}\ \bibnamefont {Kevrekidis}}, \bibinfo {author} {\bibfnamefont
  {P.}~\bibnamefont {Engels}}, \bibinfo {author} {\bibfnamefont
  {V.}~\bibnamefont {Achilleos}}, \bibinfo {author} {\bibfnamefont {D.~J.}\
  \bibnamefont {Frantzeskakis}},\ and\ \bibinfo {author} {\bibfnamefont
  {J.}~\bibnamefont {Cuevas}},\ }\bibfield  {title} {\bibinfo {title} {Beating
  dark--dark solitons in bose--einstein condensates},\ }\href@noop {}
  {\bibfield  {journal} {\bibinfo  {journal} {Journal of Physics B: Atomic,
  Molecular and Optical Physics}\ }\textbf {\bibinfo {volume} {45}},\ \bibinfo
  {pages} {115301} (\bibinfo {year} {2012})}\BibitemShut {NoStop}%
\bibitem [{\citenamefont {Danaila}\ \emph {et~al.}(2016)\citenamefont
  {Danaila}, \citenamefont {Khamehchi}, \citenamefont {Gokhroo}, \citenamefont
  {Engels},\ and\ \citenamefont {Kevrekidis}}]{danaila2016vector}%
  \BibitemOpen
  \bibfield  {author} {\bibinfo {author} {\bibfnamefont {I.}~\bibnamefont
  {Danaila}}, \bibinfo {author} {\bibfnamefont {M.~A.}\ \bibnamefont
  {Khamehchi}}, \bibinfo {author} {\bibfnamefont {V.}~\bibnamefont {Gokhroo}},
  \bibinfo {author} {\bibfnamefont {P.}~\bibnamefont {Engels}},\ and\ \bibinfo
  {author} {\bibfnamefont {P.~G.}\ \bibnamefont {Kevrekidis}},\ }\bibfield
  {title} {\bibinfo {title} {Vector dark-antidark solitary waves in
  multicomponent bose-einstein condensates},\ }\href
  {https://doi.org/10.1103/PhysRevA.94.053617} {\bibfield  {journal} {\bibinfo
  {journal} {Phys. Rev. A}\ }\textbf {\bibinfo {volume} {94}},\ \bibinfo
  {pages} {053617} (\bibinfo {year} {2016})}\BibitemShut {NoStop}%
\bibitem [{\citenamefont {Katsimiga}\ \emph {et~al.}(2020)\citenamefont
  {Katsimiga}, \citenamefont {Mistakidis}, \citenamefont {Bersano},
  \citenamefont {Ome}, \citenamefont {Mossman}, \citenamefont {Mukherjee},
  \citenamefont {Schmelcher}, \citenamefont {Engels},\ and\ \citenamefont
  {Kevrekidis}}]{Katsimiga2020}%
  \BibitemOpen
  \bibfield  {author} {\bibinfo {author} {\bibfnamefont {G.~C.}\ \bibnamefont
  {Katsimiga}}, \bibinfo {author} {\bibfnamefont {S.~I.}\ \bibnamefont
  {Mistakidis}}, \bibinfo {author} {\bibfnamefont {T.~M.}\ \bibnamefont
  {Bersano}}, \bibinfo {author} {\bibfnamefont {M.~K.~H.}\ \bibnamefont {Ome}},
  \bibinfo {author} {\bibfnamefont {S.~M.}\ \bibnamefont {Mossman}}, \bibinfo
  {author} {\bibfnamefont {K.}~\bibnamefont {Mukherjee}}, \bibinfo {author}
  {\bibfnamefont {P.}~\bibnamefont {Schmelcher}}, \bibinfo {author}
  {\bibfnamefont {P.}~\bibnamefont {Engels}},\ and\ \bibinfo {author}
  {\bibfnamefont {P.~G.}\ \bibnamefont {Kevrekidis}},\ }\bibfield  {title}
  {\bibinfo {title} {Observation and analysis of multiple dark-antidark
  solitons in two-component bose-einstein condensates},\ }\href
  {https://doi.org/10.1103/PhysRevA.102.023301} {\bibfield  {journal} {\bibinfo
   {journal} {Phys. Rev. A}\ }\textbf {\bibinfo {volume} {102}},\ \bibinfo
  {pages} {023301} (\bibinfo {year} {2020})}\BibitemShut {NoStop}%
\bibitem [{\citenamefont {Mossman}\ \emph {et~al.}(2024)\citenamefont
  {Mossman}, \citenamefont {Katsimiga}, \citenamefont {Mistakidis},
  \citenamefont {Romero-Ros}, \citenamefont {Bersano}, \citenamefont
  {Schmelcher}, \citenamefont {Kevrekidis},\ and\ \citenamefont
  {Engels}}]{mossman2024observation}%
  \BibitemOpen
  \bibfield  {author} {\bibinfo {author} {\bibfnamefont {S.~M.}\ \bibnamefont
  {Mossman}}, \bibinfo {author} {\bibfnamefont {G.~C.}\ \bibnamefont
  {Katsimiga}}, \bibinfo {author} {\bibfnamefont {S.~I.}\ \bibnamefont
  {Mistakidis}}, \bibinfo {author} {\bibfnamefont {A.}~\bibnamefont
  {Romero-Ros}}, \bibinfo {author} {\bibfnamefont {T.~M.}\ \bibnamefont
  {Bersano}}, \bibinfo {author} {\bibfnamefont {P.}~\bibnamefont {Schmelcher}},
  \bibinfo {author} {\bibfnamefont {P.~G.}\ \bibnamefont {Kevrekidis}},\ and\
  \bibinfo {author} {\bibfnamefont {P.}~\bibnamefont {Engels}},\ }\bibfield
  {title} {\bibinfo {title} {Observation of dense collisional soliton complexes
  in a two-component bose-einstein condensate},\ }\href@noop {} {\bibfield
  {journal} {\bibinfo  {journal} {Communications Physics}\ }\textbf {\bibinfo
  {volume} {7}},\ \bibinfo {pages} {163} (\bibinfo {year} {2024})}\BibitemShut
  {NoStop}%
\bibitem [{\citenamefont {Farolfi}\ \emph {et~al.}(2020)\citenamefont
  {Farolfi}, \citenamefont {Trypogeorgos}, \citenamefont {Mordini},
  \citenamefont {Lamporesi},\ and\ \citenamefont
  {Ferrari}}]{farolifi2020observation}%
  \BibitemOpen
  \bibfield  {author} {\bibinfo {author} {\bibfnamefont {A.}~\bibnamefont
  {Farolfi}}, \bibinfo {author} {\bibfnamefont {D.}~\bibnamefont
  {Trypogeorgos}}, \bibinfo {author} {\bibfnamefont {C.}~\bibnamefont
  {Mordini}}, \bibinfo {author} {\bibfnamefont {G.}~\bibnamefont {Lamporesi}},\
  and\ \bibinfo {author} {\bibfnamefont {G.}~\bibnamefont {Ferrari}},\
  }\bibfield  {title} {\bibinfo {title} {Observation of magnetic solitons in
  two-component bose-einstein condensates},\ }\href
  {https://doi.org/10.1103/PhysRevLett.125.030401} {\bibfield  {journal}
  {\bibinfo  {journal} {Phys. Rev. Lett.}\ }\textbf {\bibinfo {volume} {125}},\
  \bibinfo {pages} {030401} (\bibinfo {year} {2020})}\BibitemShut {NoStop}%
\bibitem [{\citenamefont {Chai}\ \emph {et~al.}(2020)\citenamefont {Chai},
  \citenamefont {Lao}, \citenamefont {Fujimoto}, \citenamefont {Hamazaki},
  \citenamefont {Ueda},\ and\ \citenamefont {Raman}}]{chai2020magnetic}%
  \BibitemOpen
  \bibfield  {author} {\bibinfo {author} {\bibfnamefont {X.}~\bibnamefont
  {Chai}}, \bibinfo {author} {\bibfnamefont {D.}~\bibnamefont {Lao}}, \bibinfo
  {author} {\bibfnamefont {K.}~\bibnamefont {Fujimoto}}, \bibinfo {author}
  {\bibfnamefont {R.}~\bibnamefont {Hamazaki}}, \bibinfo {author}
  {\bibfnamefont {M.}~\bibnamefont {Ueda}},\ and\ \bibinfo {author}
  {\bibfnamefont {C.}~\bibnamefont {Raman}},\ }\bibfield  {title} {\bibinfo
  {title} {Magnetic solitons in a spin-1 bose-einstein condensate},\ }\href
  {https://doi.org/10.1103/PhysRevLett.125.030402} {\bibfield  {journal}
  {\bibinfo  {journal} {Phys. Rev. Lett.}\ }\textbf {\bibinfo {volume} {125}},\
  \bibinfo {pages} {030402} (\bibinfo {year} {2020})}\BibitemShut {NoStop}%
\bibitem [{\citenamefont {Qu}\ \emph {et~al.}(2016)\citenamefont {Qu},
  \citenamefont {Pitaevskii},\ and\ \citenamefont
  {Stringari}}]{qu2016magnetic}%
  \BibitemOpen
  \bibfield  {author} {\bibinfo {author} {\bibfnamefont {C.}~\bibnamefont
  {Qu}}, \bibinfo {author} {\bibfnamefont {L.~P.}\ \bibnamefont {Pitaevskii}},\
  and\ \bibinfo {author} {\bibfnamefont {S.}~\bibnamefont {Stringari}},\
  }\bibfield  {title} {\bibinfo {title} {Magnetic solitons in a binary
  bose-einstein condensate},\ }\href
  {https://doi.org/10.1103/PhysRevLett.116.160402} {\bibfield  {journal}
  {\bibinfo  {journal} {Phys. Rev. Lett.}\ }\textbf {\bibinfo {volume} {116}},\
  \bibinfo {pages} {160402} (\bibinfo {year} {2016})}\BibitemShut {NoStop}%
\bibitem [{\citenamefont {Qu}\ \emph {et~al.}(2017)\citenamefont {Qu},
  \citenamefont {Tylutki}, \citenamefont {Stringari},\ and\ \citenamefont
  {Pitaevskii}}]{qu2017magnetic}%
  \BibitemOpen
  \bibfield  {author} {\bibinfo {author} {\bibfnamefont {C.}~\bibnamefont
  {Qu}}, \bibinfo {author} {\bibfnamefont {M.}~\bibnamefont {Tylutki}},
  \bibinfo {author} {\bibfnamefont {S.}~\bibnamefont {Stringari}},\ and\
  \bibinfo {author} {\bibfnamefont {L.~P.}\ \bibnamefont {Pitaevskii}},\
  }\bibfield  {title} {\bibinfo {title} {Magnetic solitons in rabi-coupled
  bose-einstein condensates},\ }\href
  {https://doi.org/10.1103/PhysRevA.95.033614} {\bibfield  {journal} {\bibinfo
  {journal} {Phys. Rev. A}\ }\textbf {\bibinfo {volume} {95}},\ \bibinfo
  {pages} {033614} (\bibinfo {year} {2017})}\BibitemShut {NoStop}%
\bibitem [{\citenamefont {\"Ohberg}\ and\ \citenamefont
  {Santos}(2001)}]{Oehnberg2001dark}%
  \BibitemOpen
  \bibfield  {author} {\bibinfo {author} {\bibfnamefont {P.}~\bibnamefont
  {\"Ohberg}}\ and\ \bibinfo {author} {\bibfnamefont {L.}~\bibnamefont
  {Santos}},\ }\bibfield  {title} {\bibinfo {title} {Dark solitons in a
  two-component bose-einstein condensate},\ }\href
  {https://doi.org/10.1103/PhysRevLett.86.2918} {\bibfield  {journal} {\bibinfo
   {journal} {Phys. Rev. Lett.}\ }\textbf {\bibinfo {volume} {86}},\ \bibinfo
  {pages} {2918} (\bibinfo {year} {2001})}\BibitemShut {NoStop}%
\bibitem [{Note2()}]{Note2}%
  \BibitemOpen
  \bibinfo {note} {In Ref.~{\cite {Oehnberg2001dark}}, magnetic solitons are
  referred to as kink-antikink solitons.}\BibitemShut {Stop}%
\bibitem [{\citenamefont {Charalampidis}\ \emph {et~al.}(2016)\citenamefont
  {Charalampidis}, \citenamefont {Wang}, \citenamefont {Kevrekidis},
  \citenamefont {Frantzeskakis},\ and\ \citenamefont
  {Cuevas-Maraver}}]{charalampidis2016SO2}%
  \BibitemOpen
  \bibfield  {author} {\bibinfo {author} {\bibfnamefont {E.~G.}\ \bibnamefont
  {Charalampidis}}, \bibinfo {author} {\bibfnamefont {W.}~\bibnamefont {Wang}},
  \bibinfo {author} {\bibfnamefont {P.~G.}\ \bibnamefont {Kevrekidis}},
  \bibinfo {author} {\bibfnamefont {D.~J.}\ \bibnamefont {Frantzeskakis}},\
  and\ \bibinfo {author} {\bibfnamefont {J.}~\bibnamefont {Cuevas-Maraver}},\
  }\bibfield  {title} {\bibinfo {title} {S{O}(2)-induced breathing patterns in
  multicomponent bose-einstein condensates},\ }\href
  {https://doi.org/10.1103/PhysRevA.93.063623} {\bibfield  {journal} {\bibinfo
  {journal} {Phys. Rev. A}\ }\textbf {\bibinfo {volume} {93}},\ \bibinfo
  {pages} {063623} (\bibinfo {year} {2016})}\BibitemShut {NoStop}%
\bibitem [{\citenamefont {Wang}\ \emph {et~al.}(2021)\citenamefont {Wang},
  \citenamefont {Zhao}, \citenamefont {Charalampidis},\ and\ \citenamefont
  {Kevrekidis}}]{wang2021dark}%
  \BibitemOpen
  \bibfield  {author} {\bibinfo {author} {\bibfnamefont {W.}~\bibnamefont
  {Wang}}, \bibinfo {author} {\bibfnamefont {L.-C.}\ \bibnamefont {Zhao}},
  \bibinfo {author} {\bibfnamefont {E.~G.}\ \bibnamefont {Charalampidis}},\
  and\ \bibinfo {author} {\bibfnamefont {P.~G.}\ \bibnamefont {Kevrekidis}},\
  }\bibfield  {title} {\bibinfo {title} {Dark--dark soliton breathing patterns
  in multi-component bose--einstein condensates},\ }\href@noop {} {\bibfield
  {journal} {\bibinfo  {journal} {Journal of Physics B: Atomic, Molecular and
  Optical Physics}\ }\textbf {\bibinfo {volume} {54}},\ \bibinfo {pages}
  {055301} (\bibinfo {year} {2021})}\BibitemShut {NoStop}%
\bibitem [{\citenamefont {Ho}\ and\ \citenamefont
  {Shenoy}(1996)}]{ho1996binary}%
  \BibitemOpen
  \bibfield  {author} {\bibinfo {author} {\bibfnamefont {T.-L.}\ \bibnamefont
  {Ho}}\ and\ \bibinfo {author} {\bibfnamefont {V.}~\bibnamefont {Shenoy}},\
  }\bibfield  {title} {\bibinfo {title} {Binary mixtures of bose condensates of
  alkali atoms},\ }\href@noop {} {\bibfield  {journal} {\bibinfo  {journal}
  {Physical review letters}\ }\textbf {\bibinfo {volume} {77}},\ \bibinfo
  {pages} {3276} (\bibinfo {year} {1996})}\BibitemShut {NoStop}%
\bibitem [{\citenamefont {Paw{\l}owski}\ and\ \citenamefont
  {Rz\k{a}\.{z}ewski}(2015)}]{pawlowski2015dipolar}%
  \BibitemOpen
  \bibfield  {author} {\bibinfo {author} {\bibfnamefont {K.}~\bibnamefont
  {Paw{\l}owski}}\ and\ \bibinfo {author} {\bibfnamefont {K.}~\bibnamefont
  {Rz\k{a}\.{z}ewski}},\ }\bibfield  {title} {\bibinfo {title} {Dipolar dark
  solitons},\ }\href@noop {} {\bibfield  {journal} {\bibinfo  {journal} {New
  Journal of Physics}\ }\textbf {\bibinfo {volume} {17}},\ \bibinfo {pages}
  {105006} (\bibinfo {year} {2015})}\BibitemShut {NoStop}%
\bibitem [{\citenamefont {Bland}\ \emph {et~al.}(2017)\citenamefont {Bland},
  \citenamefont {Paw{\l}owski}, \citenamefont {Edmonds}, \citenamefont
  {Rz{\k{a}}{\.z}ewski},\ and\ \citenamefont {Parker}}]{bland2017interaction}%
  \BibitemOpen
  \bibfield  {author} {\bibinfo {author} {\bibfnamefont {T.}~\bibnamefont
  {Bland}}, \bibinfo {author} {\bibfnamefont {K.}~\bibnamefont {Paw{\l}owski}},
  \bibinfo {author} {\bibfnamefont {M.}~\bibnamefont {Edmonds}}, \bibinfo
  {author} {\bibfnamefont {K.}~\bibnamefont {Rz{\k{a}}{\.z}ewski}},\ and\
  \bibinfo {author} {\bibfnamefont {N.}~\bibnamefont {Parker}},\ }\bibfield
  {title} {\bibinfo {title} {Interaction-sensitive oscillations of dark
  solitons in trapped dipolar condensates},\ }\href@noop {} {\bibfield
  {journal} {\bibinfo  {journal} {Physical Review A}\ }\textbf {\bibinfo
  {volume} {95}},\ \bibinfo {pages} {063622} (\bibinfo {year}
  {2017})}\BibitemShut {NoStop}%
\bibitem [{\citenamefont {Son}\ and\ \citenamefont
  {Stephanov}(2002)}]{son2002domain}%
  \BibitemOpen
  \bibfield  {author} {\bibinfo {author} {\bibfnamefont {D.~T.}\ \bibnamefont
  {Son}}\ and\ \bibinfo {author} {\bibfnamefont {M.~A.}\ \bibnamefont
  {Stephanov}},\ }\bibfield  {title} {\bibinfo {title} {Domain walls of
  relative phase in two-component bose-einstein condensates},\ }\href
  {https://doi.org/10.1103/PhysRevA.65.063621} {\bibfield  {journal} {\bibinfo
  {journal} {Phys. Rev. A}\ }\textbf {\bibinfo {volume} {65}},\ \bibinfo
  {pages} {063621} (\bibinfo {year} {2002})}\BibitemShut {NoStop}%
\bibitem [{Note3()}]{Note3}%
  \BibitemOpen
  \bibinfo {note} {For comparison, it is interesting to note that while dark
  solitons also have negative effective masses \cite {pitaevskii2016bose},
  certain dark-bright solitons have been predicted to dynamically oscillate
  between negative and positive effective mass \cite
  {zhao2020spin}.}\BibitemShut {Stop}%
\bibitem [{\citenamefont {Congy}\ \emph {et~al.}(2016)\citenamefont {Congy},
  \citenamefont {Kamchatnov},\ and\ \citenamefont
  {Pavloff}}]{congy2016dispersive}%
  \BibitemOpen
  \bibfield  {author} {\bibinfo {author} {\bibfnamefont {T.}~\bibnamefont
  {Congy}}, \bibinfo {author} {\bibfnamefont {A.}~\bibnamefont {Kamchatnov}},\
  and\ \bibinfo {author} {\bibfnamefont {N.}~\bibnamefont {Pavloff}},\
  }\bibfield  {title} {\bibinfo {title} {Dispersive hydrodynamics of nonlinear
  polarization waves in two-component bose-einstein condensates},\ }\href@noop
  {} {\bibfield  {journal} {\bibinfo  {journal} {SciPost Physics}\ }\textbf
  {\bibinfo {volume} {1}},\ \bibinfo {pages} {006} (\bibinfo {year}
  {2016})}\BibitemShut {NoStop}%
\bibitem [{\citenamefont {Pitaevskii}\ and\ \citenamefont
  {Stringari}(2016)}]{pitaevskii2016bose}%
  \BibitemOpen
  \bibfield  {author} {\bibinfo {author} {\bibfnamefont {L.}~\bibnamefont
  {Pitaevskii}}\ and\ \bibinfo {author} {\bibfnamefont {S.}~\bibnamefont
  {Stringari}},\ }\href@noop {} {\emph {\bibinfo {title} {Bose-Einstein
  condensation and superfluidity}}},\ Vol.\ \bibinfo {volume} {164}\ (\bibinfo
  {publisher} {Oxford University Press},\ \bibinfo {year} {2016})\BibitemShut
  {NoStop}%
\bibitem [{\citenamefont {Zhao}\ \emph {et~al.}(2020)\citenamefont {Zhao},
  \citenamefont {Wang}, \citenamefont {Tang}, \citenamefont {Yang},
  \citenamefont {Yang},\ and\ \citenamefont {Liu}}]{zhao2020spin}%
  \BibitemOpen
  \bibfield  {author} {\bibinfo {author} {\bibfnamefont {L.-C.}\ \bibnamefont
  {Zhao}}, \bibinfo {author} {\bibfnamefont {W.}~\bibnamefont {Wang}}, \bibinfo
  {author} {\bibfnamefont {Q.}~\bibnamefont {Tang}}, \bibinfo {author}
  {\bibfnamefont {Z.-Y.}\ \bibnamefont {Yang}}, \bibinfo {author}
  {\bibfnamefont {W.-L.}\ \bibnamefont {Yang}},\ and\ \bibinfo {author}
  {\bibfnamefont {J.}~\bibnamefont {Liu}},\ }\bibfield  {title} {\bibinfo
  {title} {Spin soliton with a negative-positive mass transition},\ }\href@noop
  {} {\bibfield  {journal} {\bibinfo  {journal} {Physical Review A}\ }\textbf
  {\bibinfo {volume} {101}},\ \bibinfo {pages} {043621} (\bibinfo {year}
  {2020})}\BibitemShut {NoStop}%
\end{thebibliography}
\end{document}